\documentclass[aps,pra,twocolumn,secnumarabic,nobalancelastpage,amsmath,amssymb,superscriptaddress]{revtex4-1}

\usepackage{graphics}      
\usepackage{graphicx}      
\usepackage{longtable}     
\usepackage{url}           
\usepackage{bm}            
\usepackage{epsf}

\newcommand{\ket}[1]{\left|#1\right\rangle}

\begin{document}
\title{Zeeman-Splitting-Assisted Quantum Logic Spectroscopy of Trapped Ions}
\author{Huanqian Loh} 
\author{Shiqian Ding}
\author{Roland Hablutzel}
\author{Gleb Maslennikov}
\affiliation{Centre for Quantum Technologies, National University of Singapore, 3 Science Dr 2, 117543, Singapore}
\author{Dzmitry Matsukevich}
\affiliation{Centre for Quantum Technologies, National University of Singapore, 3 Science Dr 2, 117543, Singapore}
\affiliation{Department of Physics, National University of Singapore, 2 Science Dr 3, 117551, Singapore}

\date{\today}

\begin{abstract}
We present a quantum logic scheme to detect atomic and molecular ions in different states of angular momentum based on their magnetic $g$-factors. The state-dependent magnetic $g$-factors mean that electronic, rotational or hyperfine states may be distinguished by their Zeeman splittings in a given magnetic field. Driving motional sidebands of a chosen Zeeman splitting enables reading out the corresponding state of angular momentum with an auxillary logic ion. As a proof-of-principle demonstration, we show that we can detect the ground electronic state of a ${^{174}}$Yb$^+$ ion using ${^{171}}$Yb$^+$ as the logic ion. Further, we can distinguish between the ${^{174}}$Yb$^+$ ion being in its ground electronic state versus the metastable ${^{2}}D_{3/2}$ state. We discuss the suitability of this scheme for the detection of rotational states in molecular ions. 
\end{abstract}


\maketitle


A single trapped ion, well-isolated from the environment, offers a model system for precision spectroscopy and metrology \cite{WinelandRMP}. Ideally, state detection of the ion proceeds in a nondestructive manner so that the experiment may be repeated multiple times to achieve the required precision. One such example is state-dependent fluorescence, where the ion appears ``bright'' by scattering light only when it is in a given state; otherwise, the ion is ``dark'' \cite{Dehmelt75, Blinov04}. For fluoresence to be a sensitive state detection method, the ion species must have a fast cycling transition, which may not always be present. To circumvent this constraint, quantum logic spectroscopy (QLS) was developed \cite{Schmidt05}. In QLS, the spectroscopy ion is co-trapped with a logic ion. Only the latter needs to possess a closed cycling transition. The internal state of the spectroscopy ion can then be transferred onto the shared motion of the two-ion system, which is in turn mapped onto the internal state of the logic ion.

In the initial demonstration of QLS \cite{Schmidt05}, the ions were initially prepared in the ground state, and excitation of ion crystal motion was achieved by driving motional sidebands of a narrow spectroscopy transition between long-lived states. Several variations of QLS have since been demonstrated. For instance, the trapped ions' motion can be driven by a laser resonant with the spectroscopy transition. The resulting photon recoil of the ions can be detected by looking at Doppler recooling times \cite{Clark10}, Doppler velocimetry \cite{Lin13} or phonon excitations \cite{Wan14}. These methods require the spectroscopy ion to have a partially closed cycling transition, which is not present in many atomic and molecular ions. Instead of multi-photon scattering, single-photon scattering can be employed to map the single-photon recoil of the spectroscopy ion onto a geometric phase \cite{Hempel13}. Decoherence-assisted spectroscopy, which is sensitive to a single-photon scattering event, can also be extended to a two-ion system as an alternative varation of QLS \cite{Clos14}. 

To retain the ability to perform spectroscopy on ions that require \textit{no} scattering events for state detection, one can use off-resonant continuous-wave or mode-locked pulsed \cite{Ding12, Leibfried12} lasers to exert a spin-dependent optical-dipole force on the ion. The force can be applied to coherently excite motion \cite{Hume11}, implement quantum phase gates \cite{MurPetit12}, or effect spin-motion coupling in combination with a microwave field \cite{Shi13, Ding14}. 

Here, we propose and experimentally demonstrate using two atomic ions a state detection technique employing state-dependent magnetic g-factors, which in turn show distinct Zeeman splittings. Such ``Zeeman-splitting-assisted quantum logic spectroscopy'' (ZS-QLS) uses lasers far-detuned from one-photon transitions to drive Raman transitions between Zeeman sublevels, hence does not require any scattering from the spectroscopy ion. Being a Raman version of the original implementation of QLS \cite{Schmidt05}, ZS-QLS uses long-lived Zeeman sublevels instead of a metastable electronic excited state. Nevertheless, ZS-QLS as a state detection method can be combined with other techniques (e.g.\ laser excitation) to perform spectroscopy in the optical domain.


To implement ZS-QLS, one begins with a Doppler-cooled logic ion (L) that sympathetically cools a spectroscopy ion (S). Together, they form a two-ion crystal and share common modes of motion. The motional modes along one direction are cooled to the ground state of motion by sideband cooling the logic ion.
\begin{figure*}[htb]
\includegraphics[width=17cm]{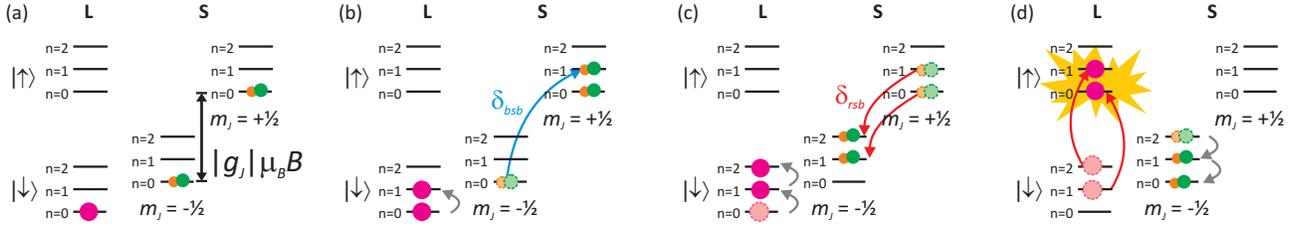}
\caption{(Color online.) Schematic of Zeeman-splitting quantum logic spectroscopy for $g_J < 0$. (a) Both the logic ion (L) and its co-trapped spectroscopy ion (S) are sideband-cooled to the ground motional state. To probe if the spectroscopy ion is in the $J = 1/2$ state regardless of its distribution amongst the Zeeman sublevels, (b) a blue sideband of the Zeeman transition is applied, followed by (c) the corresponding red sideband. A mixture of one and two phonons can then be excited. (d) The presence of phonons is detected by driving the red sideband of the logic qubit transition $|\downarrow\rangle \rightarrow |\uparrow\rangle$ and measuring the logic ion fluorescence.}
\label{fig:QLSscheme}
\end{figure*}

To probe the Zeeman splitting of the spectroscopy ion in a state of angular momentum $J$ and magnetic g-factor $g_J < 0$, a pair of Raman lasers are applied such that their frequency difference $\delta_{\mathrm{bsb}}$ is blue detuned by one normal mode frequency $\omega_n$ from the Zeeman splitting $|g_J|\mu_B B$ ($\mu_B$ is the Bohr magneton) (Fig.~\ref{fig:QLSscheme}b). For an ion initialized in the state $\ket{J, m_J}$ where the magnetic quantum number $m_J$ is less than $J$, a phonon will be added to the motion of the ion crystal with high probability. However, if the spectroscopy ion is initially in the state $|J, m_J = J\rangle$, the blue-sideband pulse would not have added any phonons. One therefore needs to apply a red-sideband Raman pulse $\delta_{\mathrm{rsb}}$ to add any phonons (Fig.~\ref{fig:QLSscheme}c) \footnote[1]{{ For an ion driven to the $|J, m_J + 1 \rangle$ state after the first Zeeman blue-sideband pulse, the subsequent red-sideband pulse can drive the ion to the $|J, m_J + 2 \rangle$ (if available) or $|J, m_J \rangle$ state, thereby removing or adding another phonon. Nevertheless, the combined action of the Zeeman blue and red sideband pulses serves to excite a phonon with high probability.}}. Applying the blue and red sideband pulses sequentially ensures that phonons can be excited regardless of the probability distribution between the Zeeman sublevels. ZS-QLS can also be applied to states of $g_J > 0$, in which case the first Zeeman blue-sideband pulse addresses all states $|J, m_J > -J\rangle$ and the red-sideband pulse ensures that $|J, m_J = -J\rangle$ is addressed.

Finally, the presence of phonons are detected with the logic ion, which fluoresces brightly when its internal state is successfully flipped upon driving the red motional sideband of the logic qubit transition (Fig.~\ref{fig:QLSscheme}d). Since the red-sideband Rabi frequencies for one- and two-phonon states are different, one has to choose a pulse time that maximizes the overall detection probability. 

If the two-photon Raman detuning is far from the Zeeman splitting of the chosen state, no phonons are added and the logic ion remains dark. Likewise, if the spectroscopy ion resides in a state of different magnetic g-factor from the chosen two-photon Raman transition, the logic ion also remains dark.

The suitability of ZS-QLS for a given atomic or molecular species depends on three main factors: I.\ the lifetime of the state whose Zeeman splitting is being probed must be long enough to resolve the motional sidebands; II.\ the spectroscopy ion must have states of different magnetic $g$-factors; III.\ the Zeeman splitting between $m_J$ sublevels in a given $J$ state should not overlap with splittings from other $J$ states. Given that a large class of molecular ions possess rotational states that are relatively long-lived (satisfying I) and have different $g$-factors (satisfying II), we now examine the applicability of ZS-QLS to molecular ions in greater detail.

In molecular ions, the magnetic $g$-factor comes from three contributions: the electron (spin and orbital) angular momentum, nuclei spin, and molecule rotation. The second and third contributions are small (on the order of $1/1840$ compared to the first contribution). In the absence of hyperfine splitting, we can focus on the first contribution.

For states of Hund's coupling case (a), the dependence of $g_J$ on the rotational level $J$ goes as \cite{Herzberg}
\begin{equation} \label{eqn:g_a}
g_J = \frac{(\Lambda + 2\Sigma)(\Lambda + \Sigma)}{J(J+1)} \, ,
\end{equation} 
where $\Lambda$ and $\Sigma$ are projections of the electron orbital and spin angular momentum on the molecular axis, respectively. While $g_J = 0$ for $^2\Pi_{1/2}$ states, the numerator of $g_J$ from Eq.~(\ref{eqn:g_a}) is 1 for $^{1}\Pi$ and 2 for $^{2}\Pi_{3/2}$, respectively. The $J$-dependence of $g_J$ does not rely on the presence of any particular molecular transitions or spectroscopic constants such as diagonal Franck-Condon overlaps, therefore any molecule with a low-lying case (a) $^1\Pi$ or $^2\Pi_{3/2}$ state such as  O$_{2}^{+}$ and SO$^{+}$ can be detected with ZS-QLS. Further, quadratic Zeeman shifts are expected to be small for reasonable table-top magnetic fields ($<200$~Hz for 5~$G$), satisfying constraint III for ZS-QLS. 

For states described by Hund's coupling case (b), the magnetic $g$-factor in the weak magnetic field is \cite{Herzberg}
\begin{eqnarray} \label{eqn:g_b}
g_{JNS} &=& \Lambda^2 \frac{J(J+1) + N(N+1) - S(S+1)}{2J(J+1)N(N+1)} \nonumber \\
&&  +\, \frac{J(J+1) + S(S+1) - N(N+1)}{J(J+1)} \, ,
\end{eqnarray}
where $S$ is the electron spin quantum number and $N = J - S$. In the strong-field regime, however, the coupling between $S$ and $N$ are weaker than their individual couplings to the applied magnetic field, thus the magnetic $g$-factor is split into two components:
\begin{equation} \label{eqn:g_b_PaschenBack}
g_N = \frac{\Lambda^2}{N(N+1)} \, ; \quad g_S = 2 \, .
\end{equation}

For the $^1\Sigma$ state, the magnetic $g$-factor is 0, so the state cannot be detected using ZS-QLS. For the $^2\Sigma$ state, $\Lambda = 0$, leaving only the second term in Eq.~(\ref{eqn:g_b}). One must therefore be careful to avoid the strong-field regime, where the $g$-factor loses its dependence on the rotational quantum number. Further, in molecular ions like CO$^+$ amd SiO$^+$ where ${^2}\Sigma$ is the ground state, requirement III is fulfilled only in weak magnetic fields ($B \lesssim 2$~$G$ for SiO$^+$ \cite{Scholl95}). For stronger magnetic fields, the quadratic Zeeman shift for low-lying rotational levels can be significant compared to the difference in Zeeman splittings between adjacent rotational levels. One exception is the $J = 1/2$ state: only comprising two Zeeman sublevels, it has no quadratic shift and $g_J = g_S = 2$. ZS-QLS can then be applied up to stronger magnetic fields ($B \lesssim 20$~$G$ for SiO$^+$), as long as other rotational states are not in the strong-field regime and possess the same magnetic $g$-factor (Eq.~(\ref{eqn:g_b_PaschenBack})).


To demonstrate the feasibility of ZS-QLS for state detection, we implemented the proposed scheme with $^{171}$Yb$^+$ as the logic ion and $^{174}$Yb$^+$ as the spectroscopy ion. In $^{171}$Yb$^+$, states $|\downarrow\rangle \equiv |{^2}S_{1/2}, F = 0, m_F = 0\rangle$ and $|\uparrow\rangle \equiv |{^2}S_{1/2}, F = 1, m_F = 0\rangle$ are chosen as qubit states \cite{Olmschenk07}. In $^{174}$Yb$^+$, $|g_J| = 2$ for the ground state. 

Both ions are trapped in a linear RF Paul trap with secular trap frequencies $(\omega_x, \omega_y, \omega_z) = 2\pi~(0.91, 0.97, 0.79)$~MHz for a single $^{171}$Yb$^+$ ion. $^{171}$Yb$^+$ is first loaded by performing resonance-enhanced two-photon ionization on Yb atoms emitted from an oven. The first photon comes from 399~nm light resonant with the $^{1}S{_0}-{^1}P{_1}$ transition in neutral $^{171}$Yb whereas the second photon for ionization comes from 369~nm light. Once loaded, the ion is Doppler-cooled by the 369~nm laser (3~W/cm$^2$) that is red-detuned from the ${^2}S{_{1/2}}-{^2}P{_{1/2}}$ transition and a 935~nm repumper laser (25~W/cm$^2$) tuned to the ${^2}D{_{3/2}}-{^3}[3/2]{_{1/2}}$ transition. $^{174}$Yb$^+$ is subsequently loaded by tuning the 399~nm laser to the neutral $^{174}$Yb resonance for photoionization. Since the 369 nm and 935~nm lasers are 2.4~GHz and 2.7~GHz red-detuned from their respective transitions in $^{174}$Yb$^+$, the $^{174}$Yb$^+$ ion is not efficiently Doppler-cooled and is instead sympathetically cooled by the $^{171}$Yb$^+$ ion. When the two ions form a crystal, they are displaced along the $\hat{z}$ direction and share the axial in-phase and out-of-phase normal modes of motion with frequencies $(\omega_{ip}, \omega_{op}) = 2\pi(0.78, 1.37)$~MHz.

The Raman lasers in this experiment come from a mode-locked Ti:Sapphire laser (pulse duration 3~ps, repetition rate 76~MHz) that is frequency-doubled using a lithium triborate crystal to give 200~mW at 377.2~nm. The doubled light is split into two beams that are independently frequency-shifted using two separate acousto-optical modulators (AOMs). The two Raman beams are sent from orthogonal directions and interfere at the trap center to form a running-wave optical lattice along $\hat{z}$. Depending on the frequency difference between the two AOMs, the Raman lasers can drive the qubit carrier transition in $^{171}$Yb$^+$, its motional sidebands, or the $^{174}$Yb$^+$ Zeeman sublevel motional sidebands. The use of a frequency comb provides the added advantage of being able to drive all the above Raman transitions with a single laser \cite{Hayes10}. The polarizations of the Raman lasers are set to be linear and oriented such that all three components $\pi, \sigma^+, \sigma^-$ relative to the quantization axis are present. This allows us to drive the $\Delta m_J = 0$ qubit transition for efficient sideband cooling as well as the $\Delta m_J = \pm 1$ transitions in $^{174}$Yb$^+$.

\begin{figure}[tb]
\includegraphics[width=8cm]{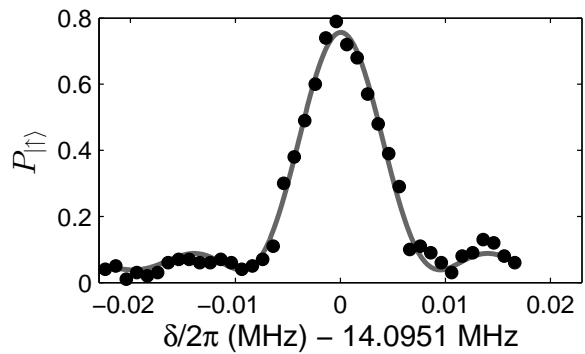}
\caption{Probability of $^{171}$Yb$^+$ occupying the bright $|\uparrow\rangle$ state as a function of the average Raman laser detuning from the ${^2}S{_{1/2}}$ Zeeman splitting in $^{174}$Yb$^+$.}
\label{fig:zeemansimulfreqscan}
\end{figure}

To begin ZS-QLS, the Raman lasers are tuned to the in-phase and out-of-phase normal mode red sidebands of the $^{171}$Yb$^+$ qubit transition for sideband cooling. One hundred sideband-cooling cycles are applied, leaving the two-ion crystal with an average phonon number of $\bar{n} = 0.029(15)$ for the out-of-phase normal mode. With $^{171}$Yb$^+$ initialized in $|\downarrow\rangle$ \cite{Olmschenk07}, the Raman lasers then apply a blue-sideband pulse ($\delta_{\mathrm{bsb}} = |g_J|\mu_B B + \omega_{op}$) followed by a red-sideband pulse ($\delta_{\mathrm{rsb}} = |g_J|\mu_B B - \omega_{op}$), each of $\pi$-pulse time 100~$\mu$s. Here, the magnetic field $B = 5.0$~$G$ is applied at 45$^\circ$ from $\hat{z}$. The out-of-phase normal mode is chosen because it exhibits a low heating rate of 2.9(4)~phonons/s compared to the in-phase mode heating rate of 70(3)~phonons/s. Finally, the presence of phonons is detected by tuning the Raman laser to the red sideband of the $^{171}$Yb$^+$ qubit transition and applying the standard state detection technique to the logic ion \cite{Olmschenk07}. Figure~\ref{fig:zeemansimulfreqscan} shows a scan of the Raman laser detuning about the Zeeman splitting. The frequency difference between the two Raman pulses is fixed to be $2\omega_{op}$ while the average frequency $\delta$ is varied. The spectrum is fit to a Rabi lineshape with a peak position at $|g_J| \mu_B B = $ 14.0951(1)~MHz. The inferred $|g_J|$ of 2 is consistent with the $^{174}$Yb$^+$ ion occupying its ground electronic state.

\begin{figure}[tb]
\includegraphics[width=8cm]{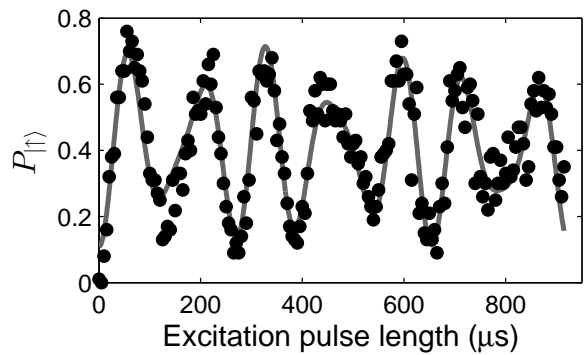}
\caption{Probability of $^{171}$Yb$^+$ occupying the bright $|\uparrow\rangle$ state after applying the Zeeman blue and red sidebands, as a function of the $^{171}$Yb$^+$ qubit red sideband pulse duration. The data is fit to a sum of Rabi oscillations arising from one phonon and two phonons that are populated with the probabilities 0.70(3) and 0.30(3), respectively.}
\label{fig:twophonondelayscan}
\end{figure}

The initial population distribution between the two $|m_J\rangle$ sublevels can also be mapped onto the probabilities of populating one~phonon versus two~phonons. The Rabi oscillations detected on the qubit red sideband are then a superposition of those with Rabi frequencies $\Omega\eta$ and $\Omega\eta\sqrt{2}$ respectively, where $\Omega/(2\pi) = 50.67(7)~$kHz refers to the qubit carrier Rabi frequency and $\eta = 0.15$ is the Lamb-Dicke parameter. The pulse durations of both the Zeeman blue and red sidebands are set to be $\pi/(\Omega\eta)$. Hence, the probability of exciting a second phonon given one phonon already being present is $p_{-}\sin^2(\sqrt{2}\pi/2)$, where $p_{-}$ is the initial probability of occupying the $\ket{m_J = -1/2}$ state. The probability of exciting only one phonon in the Zeeman scheme is $p_{-}(1-\sin^2(\sqrt{2}\pi/2)) + p_{+}$, where $p_{+}$ is the initial probability of occupying the the $\ket{m_J = +1/2}$ state. Figure~\ref{fig:twophonondelayscan} shows a scan of the pulse duration of the qubit red sideband. According to the fit, the two-phonon and one-phonon modes are populated with probabilities 0.30(3) and 0.70(3). One can therefore infer the initial probabilities $p_{-}$ and $p_{+}$ to be 0.47(5) and 0.53(4), respectively, in accordance with the fact that no particular state preparation of $^{174}$Yb$^+$ was carried out.

\begin{figure}[tb]
\includegraphics[width=8cm]{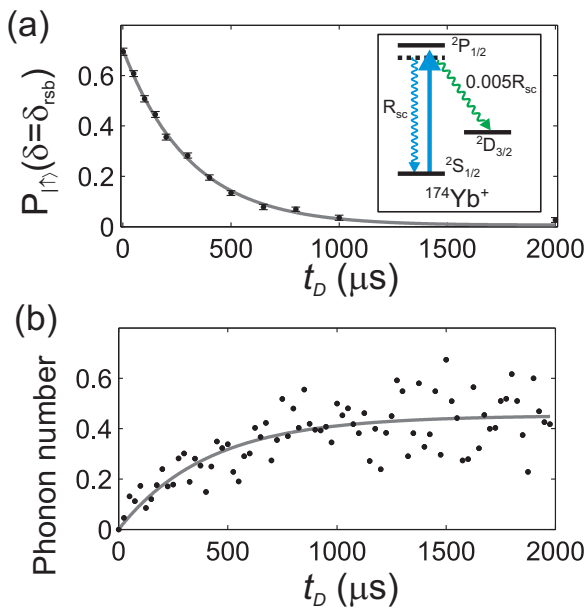}
\caption{(Color online.) Optical pumping of ${^{174}}$Yb$^+$ into the ${^{2}}D_{3/2}$ state. (a) The loss of population in the $^{2}S_{1/2}$ state can be probed by measuring the on-resonance probability of the logic ion in the bright state $P_{|\uparrow\rangle} (\delta = \delta_{rsb})$ as a function of the $^{2}D_{3/2}$ state optical pumping time $t_D$. The data is fit to an exponential decay with a decay time constant of 312(8)~$\mu$s. (Inset) Energy levels and branching ratios in $^{174}$Yb$^+$. (b) Phonon number versus optical pumping time $t_D$ for the out-of-phase normal mode. The data is fit to the functional form $n = n_0 (1 - e^{-t_D/\tau}).$}
\label{fig:zeemanoptpump}
\end{figure}

Besides populating the ground state, the $^{174}$Yb$^+$ ion can be optically pumped into its metastable $^{2}D_{3/2}$ state where $|g_J| \neq 2$, by tuning the 935~nm repumper laser far off resonance while scattering off a 369~nm laser detuned 300~MHz from its ${^2}S{_{1/2}}-{^2}P{_{1/2}}$ transition. The branching ratio for decay from the $^{2}P_{1/2}$ state is 1:0.005 (Fig.~\ref{fig:zeemanoptpump}a inset) \cite{Olmschenk07}. The ${^2}D_{3/2}$ optical pumping sequence is followed by at least two cycles of sideband cooling to remove phonons added from photon-recoil heating, so that the ions remain in the motional ground state before applying the Zeeman sideband pulses. We then take the height of the peak as recorded in Fig.~\ref{fig:zeemansimulfreqscan} for different optical pumping times $t_D$.  Figure~\ref{fig:zeemanoptpump}a shows an exponential decay with a fit time constant of 312(8)~$\mu$s, which corresponds to a decay of the $^{2}S_{1/2}$ population as the $^{174}$Yb$^+$ probability of occupying the ${^2}D_{3/2}$ state increases. The natural lifetime of the $^{2}D_{3/2}$ state is 52.7~ms and exerts a negligible effect on the measured decay.

As a check on the ${^2}D_{3/2}$ state pumping of $^{174}$Yb$^+$, we measure the photon-recoil heating induced by the 369~nm optical pumping laser on the two-ion crystal. We also start with the ions sideband-cooled to the motional ground state. Photon recoil of the $^{174}$Yb$^+$ ion adds phonons, which are detected by comparing the ratios of the blue to red sideband amplitudes of the qubit transition \cite{Wan14}. The phonon number eventually saturates when the $^{174}$Yb$^+$ ion is pumped into the $^{2}D_{3/2}$ state. The time constant for optical pumping as measured with the out-of-phase normal mode sidebands is 400(60)~$\mu$s (Fig.~\ref{fig:zeemanoptpump}b), which agrees to within 1.5~$\sigma$ of the exponential decay time measured using ZS-QLS. 


In summary, we have proposed and demonstrated a quantum logic scheme for detecting the states of dark ions based on the state-dependent magnetic $g$-factor. ZS-QLS is particularly well-suited for detecting rotational states in molecular ions. Given the lack of closed cycling transitions in molecular ions, existing popular methods for rovibrational state detection include resonance-enhanced multiphoton dissociation \cite{Roth06, Hojbjerre09, Rellergert13, Ni14, Seck14} and laser-induced charge transfer \cite{Schlemmer99, Tong10}, both of which are destructive to the ion species. On the contrary, ZS-QLS is nondestructive, allowing for repetitive experiments with a single molecular ion that forms an ideal system for precision measurement. Since ZS-QLS does not rely on special molecular constants, it can pave the way for quantum control in a wide range of species.

This research was supported by the National Research Foundation and the Ministry of Education of Singapore.

%


\begin{thebibliography}{27}
\bibitem [{Notea()}]{Notea}%
  \BibitemOpen
  \bibinfo {note} {{ For an ion driven to the $|J, m_J + 1 \delimiter "526930B
  $ state after the first Zeeman blue-sideband pulse, the subsequent
  red-sideband pulse can drive the ion to the $|J, m_J + 2 \delimiter "526930B
  $ (if available) or $|J, m_J \delimiter "526930B $ state, thereby removing or
  adding another phonon. Nevertheless, the combined action of the Zeeman blue
  and red sideband pulses serves to excite a phonon with high
  probability.}}\BibitemShut {Stop}%
\end{thebibliography}
\end{document}